\def \be {\begin{equation}}

\def \half{\frac{1}{2}}
\def \ox {\otimes}

\def \real {{\bf R}}
\def \rf  {(\ref}
\def \eqn {equation}
\def \sln {solution}

\def \tfn {transformation}
\def \mi {matri}
\def\be{\begin{equation}}
\def\ee{\end{equation}}
\def\lbl{\label}

\documentstyle[12pt]{article}

\author {Ladislav   Hlavat\'{y}
\\ Department  of  Physics,
\\ Faculty  of  Nuclear  Sciences  and Physical Engineering,
\\ B\v{r}ehov\'{a} 7, 115 19, Prague 1, Czech Republic.
\\ E-mail: hlavaty@br.fjfi.cvut.cz}

\title{Towards the Lax formulation of $SU(2)$ principal models with
nonconstant metric}

\begin{document}
\bibliographystyle{unsrt}
\maketitle
\abstract{The equations that define the Lax pairs for generalized principal
chiral models can be solved for any constant nondegenerate bilinear form on
$su(2)$. The \sln{} is dependent on one free variable that can serve as the
spectral parameter. Necessary conditions for the nonconstant metric on
$SU(2)$ that define the integrable models are given.}
\section {Formulation of generalized principal chiral models}
Principal chiral models are important example of relativistically invariant
field theory.  They are given by the action
\be I[g]=-\int d^2x \eta^{\mu\nu}L(A_\mu,A_\nu),\lbl{invaction}
\ee
where\be A_{\mu}:=-i(g^{-1}\partial_\mu g)\in{\cal L}(G), \lbl{amu}\ee
$g:\real^2\rightarrow G,\ \mu,\nu\in\{0,1\},\ \eta:=diag(1,-1)$ and $L$ is
the Killing form on the corresponding Lie algebra ${\cal L}(G)$. They are
integrable by the inverse scattering method. Their Lax pair formulation was
given in \cite{zami:relinm}.  (For Lax formulation of related $O(N)$-sigma
models see \cite{pohlm:ihs}, \cite{lund:erci}.)

An immediate generalization of the principal chiral models is obtained when
in the action \rf{invaction}) one considers a general bilinear form instead
of the Killing  \cite{soch:igpcm}.
Such a type of models were introduced e.g. as a quasiclassical limit of the
Baxter quantum XYZ model \cite{cher:relinm}.
Next step of generalization \rf{invaction}) is introducing $G-$dependent
symmetric bilinear forms into the action. In the coordinate dependent
version  the generalized principal chiral are defined by the action
\be I[g]=\int d^2x L_{ab}(g)\eta^{\mu\nu}(g^{-1}\partial_\mu g)^a
(g^{-1}\partial_\nu g)^b,
\lbl{action}\ee
where $L_{ab}(g)$ is matrix $dim G\times dim G$ defined by the $G-$
dependent bilinear form $L(g)$ as
\be L_{ab}(g):=L(g)(t_a\ox t_b), \
\ee
and $t_j$ are elements of a basis in the Lie algebra of the left-invariant
fields. It is useful to consider the bilinear form $L(g)$ as a metric on the
group manifold.
Lie products of elements of the  basis define the structure coefficients
\be [t_a,t_b]=i{f_{ab}}^ct_c
\lbl{crfort}\ee
and in the same basis we define the coordinates of the field $A$
\be iA_\nu=(g^{-1}\partial_\nu g)=(g^{-1}\partial_\nu g)^bt_b= iA_\nu^bt_b.
\ee

Varying the action \rf{action}) w.r.t. $\eta:=g^{-1}\delta g$ we obtain \eqn
s of motion for the generalized principal chiral models
\be \partial_\mu A^{\mu,a}+{\Gamma^a}_{bc}A_\mu^bA^{\mu,c}=0
\lbl{eqnsmot}\ee
where
\be {\Gamma^a}_{bc}:={S^a}_{bc}+{\gamma^a}_{bc},\lbl{Gamma}\ee
and ${S^a}_{bc}$ is the so called flat connection given by the structure
coefficients
\be {S^a}_{bc}:=-\half({F^a}_{bc}+{F^a}_{cb}),\ {F^a}_{bc}:=(L^{-1})^{ap}
{f_{pb}}^qL_{qc}
\lbl{sdef}\ee
while ${\gamma^a}_{bc}$ is the Christoffel symbol for the metric $L_{ab}(g)$
on the group manifold
\be {\gamma^a}_{bc}:=\half (L^{-1})^{ad}(U_b L_{cd}+U_c L_{bd}- U_d L_{bc}).
\lbl{christo}\ee
The vector fields $U_a$ in \rf{christo}) are defined in the local group
coordinates $\theta_i$ as
\be U_a:=U_a^i(\theta)\frac{\partial}{\partial\theta_i} \ee
where the matrix $U$ is inverse to the matrix $V$ of vielbein coordinates
\be U_a^i(\theta):=(V^{-1})_a^i(\theta),\ \
V_i^a(\theta):=-i(g^{-1}\frac{\partial g}{\partial\theta_i} )^a.\ee
Note that the connection \rf{Gamma}) is symmetric in the lower indices
\be {\Gamma^a}_{bc}={\Gamma^a}_{cb}. \ee
\section{The Lax pairs}% for general SU(2) model}
In the paper \cite{soch:igpcm},  ansatz for the Lax formulation of the
generalized chiral models was taken in the form
\be [i\partial_0+ P_{ab} A_0^b t_a+ Q_{ab} A_1^b t_a\ ,\
     i\partial_1+ P_{ab} A_1^b t_a+ Q_{ab} A_0^b t_a]=0
\lbl{lpansatz}\ee
where $P,Q$ are two auxiliary $dim G\times dim G$ \mi ces.
The ansatz \rf{lpansatz}) is a generalization of the Lax pair for $L$ equal
to the Killing form $\L_{ab}=Tr(t_at_b)$ (where $Q$ and $P$ are multiples of
the unit matrix) and for the anisotropic $SU(2)$ model where
$L_{ab}:=L_a\delta_{ab}$ (no summation) with $L_a=const.,\ L_1=L_2\neq L_3$.
A less general ansatz used in \cite{Bord:Leeds} leads just to the symmetric
spaces, i.e.$\L_{ab}:=\delta_{ab}$ for $SU(2)$.
Necessary conditions that %guarantee that
the operators in \rf{lpansatz}) form the Lax pair for the \eqn s of motion
\rf{eqnsmot}) are
\be P_{ab} {f_{pq}}^b + (P_{bp}P_{cq}-Q_{bp}Q_{cq}) {f_{bc}}^a=0,
\lbl{condN}\ee
\be \half{f_{cd}}^a(P_{cp} Q_{dq}+P_{cq} Q_{dp})
= Q_{ab} {\Gamma^b}_{pq}.
\lbl{condNS}\ee
If $Q$ is invertible, that we shall assume in the following, then these
conditions are also sufficient.
Note that the first condition is independent of the bilinear form $L$ so
that one can start with solving the \eqn{} \rf{condN}) and then look for the
bilinear forms $L$ that admit solution of the equation \rf{condNS}).

\subsection{Solution of the equation \rf{condN})}
The structure coefficients for $su(2)$ can be chosen in terms of the totally
antisymetric Levi-Civita tensor
\be {f_{ab}}^c=\epsilon_{abc}. \lbl{feps}\ee
In this case the equation \rf{condN}) can be rewritten to the form
\be (Adj\, Q)_{ab}=(Adj\, P)_{ab}+P_{ba}
\lbl{soch34a}\ee
and solved by
\be Q=\pm(P^T+Adj\, P)^{-1}/\sqrt{det(P^T+Adj\, P)}.
\lbl{qsol}\ee
where  $P^T$ is the transpose of $P$ and elements $(Adj\, N)_{ab}$ of the
adjoint \mi x to $N$ are obtained as  determinants of matrix $N$ with
dropped $b-$th row and $a-$th column mutiplied by $(-)^{a+b}$.
If we assume that $Q$ is invertible then the solution \rf{qsol}) is unique
up to the sign.

Inserting \rf{qsol}) into the conditions \rf{condNS}) that remain to be
solved we obtain rather complicated set of %algebraic equations for elements
of $P$
\eqn s of the form
\be{G^b}_{pq}(P):=\half
R_{ba}\epsilon_{cda}(P_{cp}R^{-1}_{dq}+P_{cq}R^{-1}_{dp} )
={\Gamma^b}_{pq}(L).\lbl{mastereq}\ee
where $R=P^T+Adj\, P$.
As we shall see, the fact that the left--hand side is expressed only in
terms of elements of the matrix $P$ while the right--hand side only in terms
of elements of the metric $L$ imposes restrictions on the metric for which
the above given form of Lax pair exists.
On the other hand, solvability of these  equations for L imposes conditions
for $P$ independent of $L$.
\subsection{Conditions for the metric}

As the \mi x $L_{ab}$ is symmetric we can diagonalize it  by orthogonal
transformations and the structure coefficients \rf{feps}) remain invariant.
It means that for the $SU(2)$ models we can assume without loss of
generality that $L$ is diagonal
\be L=diag(L_1(g),L_2(g),L_3(g)).
\lbl{diagL}\ee
The right--hand sides of \rf{mastereq}), i.e. elements of the connection
$\Gamma(L)$ have rather special and simple form in this case.
\be {\Gamma^a}_{bc}={\Gamma^a}_{cb}=\frac{L_b-L_c}{2L_a},\ {\rm for\ }a\neq
b, a\neq c, c\neq b.
\lbl{gam1}\ee
\be {\Gamma^a}_{bb}=-\frac{U_a L_b}{2L_a},\ {\rm for\ }a\neq b, \ {\rm no\
sums}.
\lbl{gam3}\ee
\be {\Gamma^a}_{ab}={\Gamma^a}_{ba}=\frac{U_b L_a}{2L_a},\ {\rm no\ sums}
\lbl{gam2}\ee

On the other hand, the left--hand sides of \rf{mastereq}) satisfy two
important identities, namely
\be {G^a}_{ab}=0,\ {G^a}_{bb}=0\lbl{gident}\ee
that hold for arbitrary $P$.
Indeed, using the antisymmetry of Levi-Civita symbol $\epsilon$ and
\rf{mastereq}) one gets
\[{G^b}_{bq}=\half R_{ba}\epsilon_{cda}P_{cb}R^{-1}_{dq}
= \half (Adj\,P)_{ba}\epsilon_{cda}P_{cb}R^{-1}_{dq}
= \delta_{ca}\epsilon_{cda}R^{-1}_{dq}\det P=0\]
Proof of the second identity in \rf{gident}) can be done e.g. by expressing
both inverse and adjoint \mi x via Levi-Civita symbol $\epsilon$ but it is
very tedious. The easiest way is to check it by computer.

From \rf{gident}b), \rf{mastereq}) and \rf{gam3}) we get
\be U_1(L_1-L_2-L_3)=0 {\rm \ and\ cyclic\ permutations\ of}\ (1,2,3) \ee
wherefrom we find that
\be L_1(g)=f_2(g)+f_3(g){\rm \ and\ cyclic\ permutations\ of}\ (1,2,3) \ee
where the functions $f_1,f_2,f_3$ are invariant with respect to the fields
$U_1,U_2,U_3$, respectively.
From \rf{gident}a), \rf{mastereq}) and \rf{gam2})  we get
\be \sum_{b=1}^3 U_a \log L_b=0\ \forall a\ =>\ \det L=L_1L_2L_3=const. \ee
because the the fields $U_1,U_2,U_3$ form a basis in $T_gG$.

\subsection{Conditions for the matrix $P$}
Comparing \rf{gam2}) and \rf{gam3}) we immediately see that the following
identities hold for arbitrary diagonal metric and $a\neq b$
\be L_a{\Gamma^a}_{bb}+L_b{\Gamma^b}_{ba}=0,\ {\rm no\ sums} \ee
Using the equation \rf{mastereq}), we can replace ${\Gamma^a}_{bc}$ by
${G^a}_{bc}$ and obtain a set of six linear equations for $L_j$. Their
solvability then yields three algebraic equations for elements of the matrix
$P$
\be {G^a}_{bb}{G^b}_{aa}={G^b}_{ba}{G^a}_{ba},\ {\rm no\ sums}.
\lbl{gamgam}\ee

Another set of conditions for $G^a_{bc}$ can be obtained from \rf{mastereq})
for $a\neq b, a\neq c, c\neq b$ because in these cases we get three linear
equations for $L_j$ \be {2L_a}{G^a}_{cb}={L_b-L_c}\ {\rm no\ sums}\ee
due to \rf{gam1}).
The solvability condition for these equations reads
\be G^1_{23}G^2_{13}G^3_{12}+G^1_{23}+G^2_{13}+G^3_{12}=0.
\lbl{detproL}\ee

Note that the equations \rf{gamgam}) and \rf{detproL}) are independent of
$L$. Unfortunately, they are highly nonlinear in elements of $P$ and it
seems impossible to solve them without an ansatz.

\section{Solutions for general constant metric}
As it was mentioned above, without loss of generality we can assume that the
metric $L$ is diagonal.
We shall prove that we can satisfy the conditions \rf{condN}), \rf{condNS})
for any constant diagonal $L$ by  \mi ces $P,Q$ containing one free
("spectral") parameter. It means that any generalized principal $SU(2)$
model with constant metric has a Lax pair.
\subsection{Diagonal ansatz}
The most natural extensions of the results obtained e.g. in
\cite{zami:relinm} and \cite{soch:igpcm} is the diagonal ansatz for $P$
\be P=diag(P_1,P_2,P_3).\lbl{diagp}\ee
It is rather easy to check  that in this case the conditions \rf{detproL})
and \rf{gamgam}) for the matrix $P$ are satisfied identically.

Inserting \rf{diagp}) into \rf{qsol}) one can immediately see that the
matrix $Q$ is diagonal as well, namely
$Q=diag(Q_1,Q_2,Q_3)$ where
\be Q_1=\pm \sqrt{\frac{R_{22}R_{33}}{R_{11}} },
\ Q_2=\frac{R_{33}}{Q_1}, %\epsilon \sqrt{\frac{R_3R_1}{R_2} },
\ Q_3=\frac{R_{22}}{Q_1}.
%\epsilon \sqrt{\frac{R_1R_2}{R_3} },\ \epsilon=\pm 1.
\lbl{qsoln}\ee
and
\be R_{11} =P_2P_3+P_1,\  R_{22} =P_3P_1+P_2,\  R_{33} =P_1P_2+P_3.
\lbl{rdef}\ee

Inserting \rf{diagp}), \rf{qsoln}) and \rf{rdef}) into the left-hand side of
\rf{mastereq}) for $p=q$ and $p=b$ we get zero and using \rf{gam2}),
\rf{gam3}) we find that the metric must be constant in this case.

The equations \rf{mastereq}) for diagonal $P$ reduce to three nonlinear
nonhomogeneous \eqn s for $P_j$
\be E_1:=P_1P_2[\sigma_3(P_3^2-1)+P_2^2-P_1^2]
+P_3[(P_1^2+P_2^2)\sigma_3+P_2^2-P_1^2]=0,
\lbl{peqn3}\ee
\be E_2:=P_2P_3[\sigma_1(P_1^2-1)+P_3^2-P_2^2]
+P_1[(P_2^2+P_3^2)\sigma_1+P_3^2-P_2^2]=0,
\lbl{peqn1}\ee
\be E_3:=P_3P_1[\sigma_2(P_2^2-1)+P_1^2-P_3^2]
+P_2[(P_3^2+P_1^2)\sigma_2+P_1^2-P_3^2]=0.
\lbl{peqn2}\ee
These equations, where
\[\sigma_1=\frac{L_2-L_3}{L_1},\ \sigma_2=\frac{L_3-L_1}{L_2},\
\sigma_3=\frac{L_1-L_2}{L_3}, \]
are not independent because the following relation holds identically
%is that the equations
%are functionally dependent.Using
%the relations \rf{sigdef}) we find that
\be E_1P_1L_1(-L_1+L_2+L_3)+ E_2P_2L_2(-L_2+L_3+L_1)+
E_3P_3L_3(-L_3+L_1+L_2)=0
\ee
and that's why the variety of solutions of \rf{peqn3})--\rf{peqn2})  has the
dimension one. The \sln{} curves  can be written as
\[ P_1=\kappa_1\frac{\sqrt{\mu +L_2}\sqrt{\mu +L_3}}
{\sqrt{L_2}\sqrt{L_3}},\
P_2=\kappa_2\frac{\sqrt{\mu +L_3}\sqrt{\mu +L_1}}
{\sqrt{L_3}\sqrt{L_1}},\]
\be P_3=\kappa_3\frac{\sqrt{\mu +L_1}\sqrt{\mu +L_2}}
{\sqrt{L_1}\sqrt{L_2}}
\lbl{psoln} \ee
where $\mu$ is a free  parameter and
\[ \kappa_1^2=\kappa_2^2=1,\  \kappa_3= \kappa_1\kappa_2. \]
Inserting \rf{psoln}) into \rf{qsoln}) we get
\be Q_1=\omega_1\frac{\sqrt{\mu}\sqrt{\mu +L_1}}
{\sqrt{L_2}\sqrt{L_3}},\
Q_2=\omega_2\frac{\sqrt{\mu}\sqrt{\mu +L_2}}
{\sqrt{L_3}\sqrt{L_1}},\
Q_3=\omega_3\frac{\sqrt{\mu}\sqrt{\mu +L_3}}
{\sqrt{L_1}\sqrt{L_2}},
\lbl{qsolnlam} \ee
where
\[ \omega_1=\kappa_2\omega_3,\ \omega_2=\kappa_1\omega_3,
\ \omega_3^2=1. \]

The formulas \rf{diagp}), \rf{qsoln}), \rf{psoln}), and \rf{qsolnlam}) yield
the \sln{} of the \eqn s \rf{condN}), \rf{condNS}) for
\be L_{ab}=L_a\delta_{ab},\ {f_{ab}}^c=\epsilon_{abc}
\ee
and up to an eventual orthogonal \tfn{} of the algebra basis, they define
the Lax pair for the generalized principal $SU(2)$ chiral model with the
constant anisotropic metric.
It is easy to check that  for $L_1=L_2=L_3$ and $L_1=L_2\neq L_3$, the Lax
pairs coincide with the previously known cases \cite{soch:igpcm}.
\subsection{Block--diagonal ansatz}
Our next  goal is to find model with the $G-$dependent metric. As it follows
from the preceeding subsection, the only possibility is the non--diagonal
matrix $P$. On the other hand the calculations with the general matrix seem
hopelessly complicated so that we can try the block--diagonal form
\be
P  =  \left ( \begin{array}{ccc}
  p_1 & b_1 & 0  \\
  b_2 & p_2 & 0 \\
  0 & 0 & p_3
  \end{array} \right ).
\lbl{bd}\ee
However, as we shall see, this ansatz leads again to the constant metric.

Inserting\rf{bd}) into \rf{qsol}) we find that the matrix $Q$ has the same
block--diagonal form as $P$.
The conditions \rf{gamgam}) are satisfied identically for block--diagonal
$P$ while the equation \rf{detproL}) now reads
\[ 0={p_3}\,{( b_1\,p_1 +  b_2\,p_2 ) }\,
      ( {{b_1}^2} - {{b_2}^2} -{{p_1}^2} + {{p_2}^2} )\times\]
\be      \left( 2\,p_1\,p_2 -2\,b_1\,b_2 +  p_3\,
           ( 1 + {{b_1}^2}  + {{b_2}^2} + {{p_1}^2}+
              {{p_2}^2} - {{p_3}^2})  \right).
\lbl{dplbd}\ee
On the other hand, from \rf{mastereq}), \rf{gam2}) and \rf{gam3}) we find
that
\[ U_3 L_k=(-)^k L_k\,
     {p_3}( b_1\,p_1 +  b_2\,p_2 )\times\]
\be         \left( 2\,p_1\,p_2 -2\,b_1\,b_2 +
     p_3( 1 + {{b_1}^2}  + {{b_2}^2} + {{p_1}^2}+
              {{p_2}^2} - {{p_3}^2})  \right)
\lbl{ddtbd}\ee
for $k=1,2$ and $U_aL_b=0$ for
 all other combinations of indices. Comparing \rf{dplbd})
and \rf{ddtbd}) we can see that the Lax pair for nonconstant metric can
exist only if
\be  {b_1}^2 - {b_2}^2 ={p_1}^2 - {p_2}^2 \ee
Unfortunately in this case $G^3_{12}=0$ so that from \rf{mastereq}) and
\rf{gam1}) we get $L_1=L_2$ and from \rf{ddtbd}) we find that the metric
must be constant.

\section{Conclusions}
The ansatz of the form \rf{lpansatz}) for the Lax pair formulation of the
generalized principal chiral model \rf{action}) implies rather complicated
set of equations \rf{condN}), \rf{condNS}) for elements of the matrices $P$,
$Q$ and metric $L$.

For the group $SU(2)$ the matrix $Q$ can be solved in terms of $P$ and one
can
derive admissible group dependence of the metric $L$ under which the the Lax
pair for the generalized principal chiral model may exist.

It seems that there is no other way to solve the equations \rf{condN}),
\rf{condNS}) but using an ansatz. Using the diagonal form of the matrix $P$
we have found the explicit form of the Lax pair with the spectral parameter
for the general anisotropic $SU(2)$ model with constant metric and then we
have proved that the Lax pair for the nonconstant metric (if it exists in
the form \rf{lpansatz})) requires a more general form of the matrix $P$ than
the block--diagonal \rf{bd}). Unfortunately, all more general forms that
were tried (Jordan form and others) did not simplify the equations for $P$
and $L$ to a form that would be solvable.

The author gladly acknowledge assistence of the systems Reduce 3.6
\cite{red36}  and Mathematica 3.0 \cite{mathem} by the presented
calculations.
This work was done with the support of the grant No. 828/1999 of Ministry of
Education of the Czech Republic and the grant  K1019601 of Academy of
Sciences of the Czech Republic.

\end{document}